# OWC-enabled Spine and Leaf Architecture Towards Energy Efficient Data Center Networks

Abrar S. Alhazmi, Sanaa H. Mohamed , T. E. H. El-Gorashi, and Jaafar M. H. Elmirghani
School of Electronic and Electrical Engineering, University of Leeds, LS2 9JT, United Kingdom
elasal@leeds.ac.uk, s.h.h.mohamed@leeds.ac.uk, t.e.h.elgorashi@leeds.ac.uk, j.m.h.elmirghani@leeds.ac.uk

*Abstract*— Due to the emergence of new paradigms and services such as 5G/6G, IoT, and more, current deployed wired Data Center Networks (DCNs) are not meeting the required performance metrics due to their limited reconfigurability, scalability, and throughput. To that end, wireless DCNs using technologies such as Optical Wireless Communication (OWC) have become viable and cost-effective solutions as they offer higher capacity, better energy efficiency, and better scalability. This paper proposes an OWC-based spine and leaf DCNs where the leaf switches are enabled with OWC transceivers, and the spine switches are replaced by Access Points (APs) in the ceiling connected to a backbone network. The APs are interconnected through a Passive Optical Network (PON) that also connects the architecture with upper network layers. An Infrared (IR) OWC system that employs Wavelength Division Multiplexing (WDM) is proposed to enhance the DCN downlink communication. The simulation (i.e., channel modeling) results show that our proposed data center links achieve good data rates in the data center up to 15 Gbps. For the PON, Arrayed Waveguide Grating Routers (AWGRs) that enable WDM are proposed to connect the APs. We evaluate the performance of the considered architecture in term of its power efficiency compared to traditional spine and leaf data centers. The results show that the OWC-enabled DCN reduces the power consumption by 42% compared to traditional the spine and leaf architecture.

I. INTRODUCTION

Data centers are experiencing major traffic explosions as a result of the emerging cloud and edge computing paradigms, 5G services and expected 6G services, Internet-of-Things (IoT) applications, big data applications, and more [1]. This has resulted in the setting up of rigorous standards for Data Center Networks (DCNs) in terms of their connectivity, throughput, reconfigurability, latency, and scalability. To meet these standards, technological and structural developments in the DCN design are therefore needed. Wired DCNs with copper and optical fiber cables used for intra-cluster and inter-cluster communication have received significant attention [2], [3]. Despite the benefits of wired DCNs, the hierarchical nature of the deployment of their topologies creates complications in terms of space requirement, heat removal, cabling management and reconfiguration for future upgrades. Generally, bandwidth overload, flexibility and scalability are common problems with traditional fiber/cable-connected DCNs.

Accordingly, incorporating Optical Wireless Communication (OWC) technologies into DCNs partially or fully is a viable and cost-effective remedy to the concerns stated above. Hence, a technology change has already been initiated to move from wired cable and fiber connections [4] – [5] to wireless connections in order to meet the requirements of the dynamic demands in next-generation DCNs [6]. This is particularly important given the fact that the move towards wireless technology reduces the cooling requirements as well as potentially minimizing and simplifying the data center infrastructure where no floating floor or multiple layer ceilings are needed [7]. Adopting OWC has several merits such as offering high bandwidth, where OWC can offer data rates above 25 Gb/s [8] - [15] using for example beam power adaptation, beam angle adaptation, relay nodes and diversity receivers [24] - [28]. For example, recent works reported high data rates reaching 575 Mbps and 225 Mbps in the downlink directions [15] - [20]. Similarly, previous work illustrated that OWC in the uplink direction also achieved high data above 7.14 Gb/s [8], [21]. In [21], we extended our work in [8] and examined an Infrared (IR)-based OWC system that uses Angle Diversity Transmitters with an Angle Diversity Receiver (ADR). We considered this OWC system to be used in a proposed spine-and-leaf data center architecture for the uplinks (i.e., the links from leaf or ToR switches placed on top of racks to spine switches connected to OWC-based access points on the ceiling).

The use of Passive Optical Networks (PONs) in data centers has also been proposed as they offer various features, particularly in terms of the energy efficiency. These features include the use of mostly passive components and are cost-efficient, data rate agnostic, and with low maintenance requirements. Using Arrayed Waveguide Grating Routers (AWGR) in PONs of data centers enables Wavelength Division Multiplexing (WDM) and hence, increases the bandwidth of the data center fabric [22]. An Optical Line Terminal (OLT) switch can serve and manage the communication in these data centers [22].

In this paper, we proposes using an infrared WDM OWC system to enhance the downlink communication in spine and leaf data centers that classically consist of an upper layer of spine switches and a lower layer of leaf switches with all-to-all cabling requirement between the switches in the two layers [23]. In our proposal, the leaf switches are enabled with OWC transceivers, and the spine switches are replaced by Access Points (APs) in the ceiling. We propose connecting the APs using a PON based on AWGRs. Finally, we compare the power consumption of this design to traditional spine and leaf data center architectures and show that our proposed design improves the energy efficiency of the spine-leaf data centers due to using an OWC system and a PON. To the best of our knowledge, this is the first work that proposes the use of AWGR-based PON in an OWC-enabled spine and leaf DCN.

This rest of this paper is organized as follows: Section II provides an overview of related work. Section III presents the system model. Section IV presents the data center's downlink OWC configuration. Section V provides a performance evaluation of the downlink in the OWC-enabled DCN architecture. Section VI presents the design of the AWGR-based PON for the proposed OWC-based spine and leaf DCN. Section VII discusses the power consumption of the proposed architecture and provides a benchmark against traditional spine and leaf DCNs. Finally, Section VIII concludes the paper and provides future work directions.

## II. RELATED WORKS

It has been shown that using PON technology leads to high-performance, high-energy efficiency, high capacity, low cost, scalability and design elasticity in access networks. Therefore, using this technology in contemporary and future data centers can bring equivalent features to their DCN designs. In [27], [28], the authors proposed data center networks that employ different PON technologies while maintaining the Top-of-the-Rack (ToR) electronic switches. The PON technologies proposed included Orthogonal Frequency Division Multiplexing (OFDM) PON and Wavelength Division Multiplexing (WDM) PON [24], [25]. Five unique PON-based data center architectures were presented in [26] that offered high capacity for intra and inter rack interconnections at low cost and high energy-efficiency in future DCNs. In comparison to the existing DCN architectures, these architectures completely substitute access, aggregation and core electronic switches with distinct passive intra-rack and inter-rack interconnections, along with an OLT. The authors of [27] and [28] discussed the third distinct architecture that comprised of two AWGRs to offer complete interconnection within the PON cell that includes four PON groups (i.e., four racks) using WDM. In this architecture, various photodetectors and tunable lasers are installed in every server for identifying and transmitting the different wavelengths [27]. However, these designs use only wired fibre-based links in their DCN design and would thus incorporate cabling challenges in their deployments.

In [29], [30], a new method is proposed to network data center components using the 60 GHz frequency band. The authors suggested that wireless connections with beam-forming transceivers should be used for high data rate point-to-point links in data centers. Investigation of the entire DCN architecture was addressed in [8], [31] while primarily focusing on the simulation model or the theoretical analysis. The work in [32], [33] proposed an OWC system and reported 10 Gb/s transmission experiments without contemplating the whole network architecture. The authors considered the DCN architecture and provided an experimental evaluation. By using a number of different orthogonal coding methods such as WDM, numerous access and resource allocation techniques can be utilized to minimize the interference. In particular, WDMA has recently attracted attention, considering the support for multiple racks [34], [35]. The use of WDM also enables scalable bandwidth allocation for data centers as a response to the ever-increasing bandwidth requirements. WDMA systems use a multiplexer at the transmitter side and reverse the operation in the receiver side by adding a demultiplexer to deal with the different wavelengths in the Multiple Input Multiple Output (MIMO) system [36]. In contrast to previous works, this work is the first to combine the use of OWC-based connections between the leaf layer (i.e., ToRs switches) and the spine layer (i.e. OWC APs in the ceiling), and the AWGR-based PON to connect the APs to increase the energy efficiency of spine and leaf data centers and to ease their cabling requirements. The PON-based networks can additionally provide links between the APs and core networks through the OLT, which can provide robust access to the data center racks. This feature makes the proposed design suitable for both large-scale and edge-based distributed data center environments.

## III. SYSTEM MODEL

The traditional spine and leaf data center structure, shown in Figure 1, gained wide popularity within the industry because it offers several benefits such as high-radix switches, security, and high-capacity [37]. This structure consists of the leaf layer that connects to the storage and servers, and the spine layer. This spine layer is the core layer of the structure and is responsible for interconnecting all leaf switches. In this paper, we evaluate the downlink in our proposed spine-and-leaf data center architecture using an IR-based OWC system. We report a proof-of-concept simulation setup consisting of four racks using an appropriate software platform to model the downlink function of the system model. We use an IR transmit power in the range of a few mWs to comply with eye safety requirements [38].

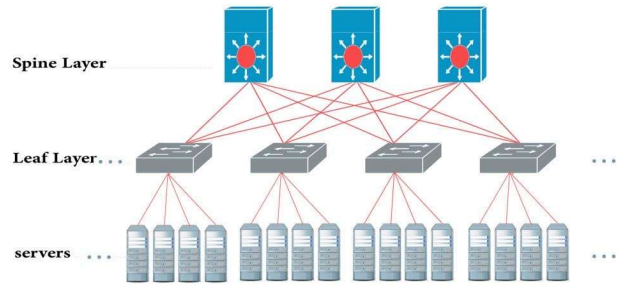

Figure 1: Spine and Leaf Traditional Architecture.

Each one of the four branches within the transceiver has one of the four wavelengths produced by a transmitter. Different wavelengths are used to avoid interference between the four IR wavelengths. The suggested wavelengths are as follows: $\lambda 1=850nm$, $\lambda 2=880nm$, $\lambda 3=900nm$, and $\lambda 4=950nm$, with 4mW transmitted power [38]. In our work, we propose using Angle Diversity Transmitter (ADT). To reduce delay spread and maximize the SNR, we select the number of branches of the ADT, the beam angles, the azimuth angles, and elevation angles given the location of the racks, in addition to the likely number and locations of receivers. The suggested ADT has four faces, and each face is directed to a specific area representing a possible receiver rack location and hence, one possible rack. We select the elevation and azimuth angles of each face to make each branch see one receiver only. For the proposed AWGR-based architecture, Figure 2, shows the different entities including, the passive components including two AWGRs and the fibers which are passive and the active elements including the OLT, ToR switches (i.e., leaf switches) and the OWC transceivers. To achieve all-to-all connection between the different racks and the OLT, we utilize four wavelengths for the PON to guide and route the traffic between the APs and with the OLT through the AWGRs. It should be noted that these wavelengths are different from the ones used for the OWC system between OWC transceivers. The APs in the ceiling are assumed to perform the wavelength conversion and optical-electrical-optical conversation needed. We keep the leaf switches to connect the servers within each rack.

The APs are connected to the AWGR so they can achieve parts of the networking functions. The AWGR provides routing to the wavelengths between their input ports and output ports. The OWC transceivers are used to remove the fiber links that originally connect the spine and the leaf layers. In our proposal, the routing/switching functions are done by the leaf switches, the AWGRs, and the OLT for the control of wavelength tuning. The performance of the OWC part of the architecture will impact the SNR and hence, the links capacity, while the choice of the PON and active networking equipment will impact the power efficiency. For the PON, we suggest using two $N \times N$ AWGRs, where $N$ is the number of racks in the data center [22]. Each AWGR is connected to the OLT port. In our design AWGRs are for inter rack communication only. For example, if rack-1 in Figure 2 wants to send data to rack-2, the process goes as follows: rack-1 will send the traffic to a specific AP on the ceiling depending on the destination. The AP is a device that has an OWC system for the interface with the racks and has a wired optical fiber interface with AWGR. The AP then sends the signal through a short fiber connection to the AWGR using the wavelength that routes to the right receiving AP. The receiving AP sends the signal to the OWC transceiver connected to it and the OWC sends the signal wirelessly to the intended destination rack, rack-2.

In what follows, we discuss the channel modeling of the downlink to complement the discussion of the uplink in our previous work in [8], [21]. Both works are to be combined to achieve the full functionality of the OWC transceivers in Figure 2.

IV. DATA CENTER DOWNLINK OWC CONFIGURATION

The size of the data center room (length × width × height) is set to 8m × 8m × 3m. The data center consists of four rows of racks [39], [40]. Every rack has its own leaf switch. Each leaf switch is connected to an OWC transceiver placed at the top of the rack as illustrated in Figure 3. It is assumed that the racks are placed with 1m spacing between them, with the same spacing assumed between the rows of each rack. A ray tracing algorithm was used to model all reflective surfaces including data center walls, ceiling, and floor [41]. Each of the surfaces in the data center are divided into small equal areas ($dA$), with a reflection coefficient ($\rho$). The simulation analyzed reflections up to the second order since any higher order reflections have negligible influence on the received signal power and delay spread [41]. Additionally, plaster walls were considered in this study. These types of surfaces have been shown to reflect signals with a Lambertian pattern [42]. The walls, ceiling and floors within the data center were therefore modelled as Lambertian reflectors with a coefficient of reflection equal to 0.8 for the ceiling and the walls and 0.3 for the floor, similar to the approach in [17], [43].

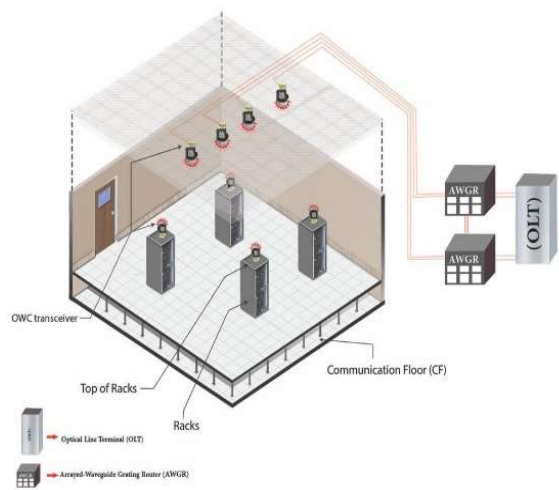

Figure 3: System model for the proposed spine and leaf data center with OWC transceivers and a PON.

The OWC system model proposed in this study comprises four ADTs deployed at four locations on the ceiling of the data center. The first transmitter is situated in the ceiling of the data center with coordinates at the corner (4m, 1m, 3m), the second transmitter is at (4m, 3m, 3m), the third transmitter is at (4m, 5m, 3m), and the fourth is located at (4m, 7m, 3m). The transmitter locations are chosen to provide good downlink connectivity for each receiver in the data center. Each ADT comprises four branches, each with two distinctive angles, the azimuth and the elevation angles, defining the orientation of each branch. The azimuth angles of the first ADT transmitter are set at 167º, 207º, 231º, and 243º respectively, with 19º, 18º, 13º and 9.5º elevation angles.

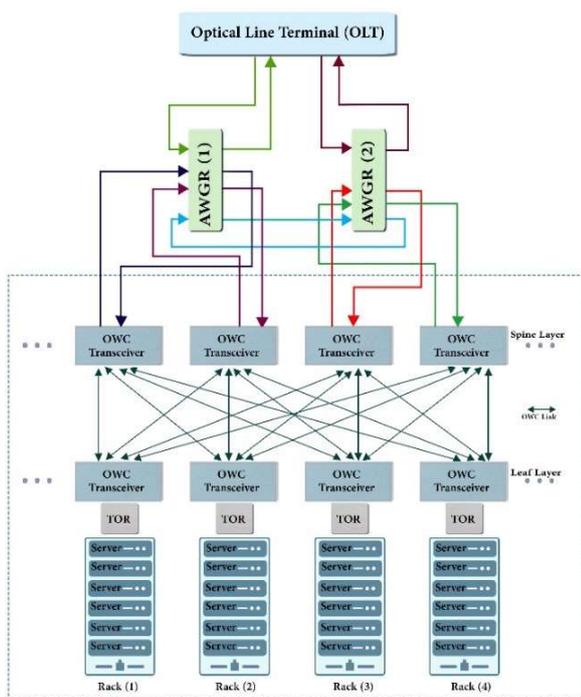

Figure 2: Proposed PON-Based Architecture for Data Center OWC.

The azimuth angles of the second ADT transmitter are 90º, 90º, 270º and 270º, and the elevation angles are 18.5º, 45º, 45º and 18.5º. The azimuth angles of the third ADT transmitter are 90º, 90º, 90º and 90º, while the elevation angles are 10º, 15º, 31º and 74º. Finally, the azimuth angles of the fourth ADT transmitter are 124º, 143º, 180º, and 216º respectively, with 11º, 16º, 20º and 16º elevation angles. The semi-angle of the beam out of each branch is set to a very small value resulting in a directed beam. The Azimuth angle, elevation angle and semi-angle values are chosen to allow each branch to be directed to one receiver only. With the above-mentioned setting, each ADT emits four beams each directed to a different receiver in a different rack, and hence, achieving the downlink function of a spine and leaf data center architecture. In this paper, two types of optical receivers are considered. The coordinates of the receivers are as follows for both types of receivers: (1.3m, 1.6m, 2m), (4m, 4m, 2m), (4m, 6.3m, 2m) and (1.3m, 5m, 2m). The first type of receiver is wide field of view (WFOV) receiver with FOV of 90º. The second type of receiver is the Angle Diversity Receiver (ADR) with FOV of 5º.

## V. PERFORMANCE EVALUATION OF THE OWC-ENABLED DCN ARCHITECTURE

This section evaluates the SNR of the links in the proposed system shown in Figure 3. The simulation tool used is MATLAB. An optical wireless channel simulator [44] was developed to generate the SNR and data rate results. The uplink results are presented in our paper [21]. The SNR should be high enough in order to enable high data rates and to reduce the probability of error. For on-off keying (OOK), the probability of error is given as:

$$P_e = Q\sqrt{SNR}, \qquad (1)$$

where Q(.) is the Gaussian function and the SNR is computed as follows [16]:

$$SNR = \frac{R^2(P_{s1} - P_{s0})^2}{\sigma_t^2}, \qquad (2)$$

where $R$ is the photo-detector responsivity, $P_{s1}$ is the power associated with logic 1, $P_{s0}$ is the power associated with logic 0, and $\sigma_t^2$ is the total noise due to the received signal and receiver noise current spectral density:

$$\sigma_t^2 = \sigma_{pr}^2 + \sigma_{bn}^2 + \sigma_{sig}^2, \qquad (3)$$

where $\sigma_{pr}^2$ is the mean square receiver preamplifier noise current, $\sigma_{bn}^2$ is the mean square background shot noise current and $\sigma_{sig}^2$ is the mean signal induced shot noise current. The achievable data rate or channel capacity is determined as [45]:

$$Channel\ capacity = B\ log_2(1 + SNR), \qquad (4)$$

where $B$ is the bandwidth of the receiver. Figure 4 shows the corresponding data rate results in Gbps for the links between different transmitters and receivers when using the ADT with two different types of receivers, namely an ADR and a WFOV receiver. It is observed that the ADR provides better data rates performance compared to the WFOV receiver.

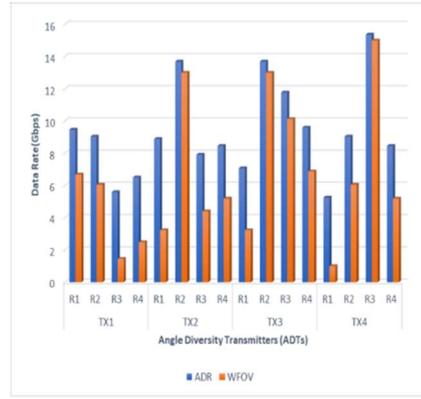

Figure 4: The OWC system Data Rates, with ADTs, ADRs and WFOV receiver Downlink.

## VI. THE PROPOSED PON-BASED NETWORK FOR DATA CENTER OPTICAL WIRELESS COMMUNICATION

The PON architecture considered is based on the proposed use of PON3 in [22], [27]. This PON, which is based on AWGRs, is used to connect the OWC-based APs. Two AWGR are proposed to provide all-to-all connections between all the APs and the OLT. Each 2 APs in a set will connect to a unique input/output port of one of the two AWGRs as shown in Figure 2. An AP input port and output port are connected to a single output port and input port of one of the AWGRs respectively, and each AWGR is connected to the OLT. To assign wavelengths for the communication between the APs and the OLT, a mixed integer linear programming (MILP) optimization problem similar to the one in [22] is formulated. The model's objective is to maximize the total number of connections supported by the AWGRs between the PON groups (i.e. the APs and the OLT). The model uses wavelength allocation constraints to ensure that each communication between the APs and the OLT and among the APs uses a wavelength and a path that complies with their functional requirements of the AWGRs, ensuring in the end all-to-all communication. A routing constraint is also used to guarantee that a wavelength is only used once to connect a pair of source and destination nodes on a physical link. Table 1 shows the obtained wavelength assignment for the proposed architecture.

**TABLE 1 Wavelength assignment for the communication between the APs and the OLT in the proposed architecture**

| | | Receiver | | | | |
|---|---|---|---|---|---|---|
| | | AP1 | AP2 | AP3 | AP4 | OLT |
| Sender | AP1 | - | $\lambda_4$ | $\lambda_1$ | $\lambda_2$ | $\lambda_3$ |
| | AP2 | $\lambda_4$ | - | $\lambda_3$ | $\lambda_4$ | $\lambda_2$ |
| | AP3 | $\lambda_1$ | $\lambda_2$ | - | $\lambda_3$ | $\lambda_4$ |
| | AP4 | $\lambda_2$ | $\lambda_3$ | $\lambda_4$ | - | $\lambda_1$ |
| | OLT | $\lambda_3$ | $\lambda_4$ | $\lambda_2$ | $\lambda_1$ | - |

## VII. POWER CONSUMPTION BENCHMARK

A second relevant metric is this work is the energy efficiency. Accordingly, in this part, we show a benchmark study to compare the power consumption of traditional spine and leaf data centers and the proposed design. The power consumption of the devices in the spine and leaf data center architecture used in this benchmark study are shown in Table 2. Furthermore, the power consumption of the devices in the PON-based OWC architecture used in this benchmark study are shown in Table 3.

**TABLE 2 Power Consumption of Different Devices in traditional spine and leaf data centers.**

| Network device | Power consumption |
|---|---|
| Spine Switch | 660 Watt [37] |
| Leaf Switch | 508 Watt [37] |
| Server Transceiver | 3 Watt [37] |

**TABLE 3 Power Consumption of Different devices in the proposed PON-based data center.**

| Network device | Power consumption |
|---|---|
| OWC transceiver | 0.4 Watt [47] |
| Leaf Switch | 508 Watt [37] |
| OLT | 480 Watt [46] |
| Server Transceiver | 3 Watt [37] |

Equation (5) calculates the power consumption of a spine and leaf network architecture:

$$P = (Ps \times Ns) + (Pl \times Nl) + (Pcs \times Ncs), \quad (5)$$

where $Ps$ is the power consumption of a spine switch, $Ns$ is the number of spine switches used in the architecture, $Pl$ is the power consumption of a leaf switch, $Nl$ is the number of used leaf switches in the architecture, $Pcs$ is the power consumption of a server's transceiver and $Ncs$ is the number of transceivers used in the architecture.

Equation (6) calculates the power consumption of our proposed PON-based data center architecture.

$$P = (Po \times No) + K + (Pl \times Nl) + (Pc \times Nc), \quad (6)$$

where $Po$ is the power consumption of the OWC transceivers, $No$ is the number of the OWC transceivers, and $K$ is the power consumption of the OLT. We considered 4 racks and 32 server per rack. For the traditional spine and leaf architecture, the number of spine switches is 4 and the number of leaf switches is 4. Our proposed OWC-enabled DCN minimizes the power consumption by 42 % compared to traditional the spine and leaf architecture as shown in Figure 5.

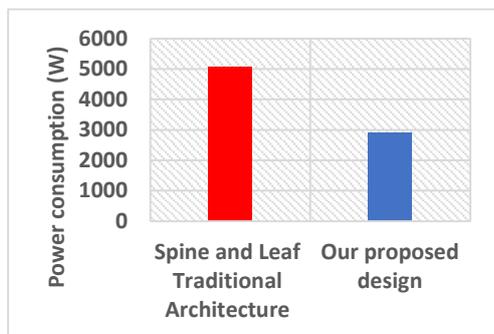

Figure 5: Power Savings of the proposed PON-based data center architecture compared to tranditional Spine and Leaf architecture.

## VIII. CONCLUSIONS

This paper proposed an OWC-based spine and leaf DCNs where the leaf switches are enabled with OWC transceivers, and the spine switches are replaced by Access Points (APs) in the ceiling connected through a WDM PON. We first evaluated the use of an IR-based OWC system with WDM to enhance DCN communication in the downlink direction. The results showed that our proposed data center links achieved a data rate up to 15 Gb/s. We then provided a benchmark study for the power consumption to compare the proposed architecture with the traditional spine and leaf architecture and our proposed architecture achieved 42% reduction in the power consumption compared to the traditional architecture. For future work and as in real-life implementations, data centers would typically contain a large number of racks (to accommodate tens or hundreds of thousands of servers (32 servers per rack is typical), we will consider modelling higher number of racks with optimum transmitters and receivers' settings.

ACKNOWLEDGMENTS

The authors would like to acknowledge funding from the Engineering and Physical Sciences Research Council (EPSRC) INTERNET (EP/H040536/1), STAR (EP/K016873/1) and TOWS (EP/S016570/1) projects. A.S.A would like to thank Taibah University in the Kingdom of Saudi Arabia for funding her PhD scholarship. All data are provided in full in the results section of this paper.